\documentclass[12pt]{article}
\usepackage[cp866]{inputenc}
\makeatletter
\renewcommand{\@oddhead}{\hfil Alexander Shatskiy $\qquad\qquad$  "Primordial black holes and asteroid danger" $\qquad\qquad$}

\begin{document}

\title{Primordial black holes and asteroid danger}
\author{Alexander Shatskiy \footnote{{\bf e-mail}: shatskiy@asc.rssi.ru}
\\
$^*$ \small{\em Astro Space Center of the P.~N.~Lebedev Physics Institute, Russian Academy of Sciences} \\
\small{\em ul. Profsoyuznaya 84/32, 117997 Moscow, Russian
Federation}}
\date{}
\maketitle

\abstract{Probability for a primordial black hole to invade the
Kuiper belt was calculated. We showed that primordial black holes
of certain masses can significantly change asteroids' orbits.
These events may result in disasters, local for our solar system
and global for the Earth (like the Tunguska meteorite). We also
estimated how often such events occur.}

\section{Introduction}
Nowadays the asteroid danger is one of the most important
scientific issues. Geophysical data indicate that in the past the
Earth was dangerously attacked by asteroids about every 190
million years rather than continuously. These repeated global
events suggest that the probability for asteroids, or meteorites,
to shoot is not constant, but depends on external (with respect to
the Solar system) factors (see~\cite{2}).

At the moment the nature of dark matter is not clear. One of the
candidates for dark matter particle is a primordial black hole
(PBH). In case the entire dark matter consists of PBHs, it is
possible to set not only the lower (Hawking) bound for their
masses, but also the upper one. The latter comes from the data on
gravitational microlensing (for example, see~\cite{Carr}).
However, the upper bound is quite uncertain and in different
estimations lies in the range ${10^{26}\div 10^{28}}$ g, i.e.
close to the Earth and Moon masses.

In this paper we obtained the probability for a PBH travelling
through the Solar system to collide (interact) with an asteroid.
This process can make collision of the Earth and some heavenly
body more probable.

To date, we are aware of several asteroids whose orbits are
threateningly close to the Earth. The asteroids that are
sufficiently big and massive are most dangerous to our planet. For
example, the Tunguska meteorite must have been around 50 m in
diameter.

As is well-known, there are three prominent asteroid belts in the
Solar system:

1. Mass of the main asteroid belt (the Phaeton belt) is about
${10^{24}}$g~${\sim 10^{-3}M_\oplus}$. Asteroids populating the
belt are 2.5---3 astronomical units (AU) away from the Sun.

2. Mass of the Kuiper belt is around ${10^{27}}$g~${\sim
M_\oplus}$. Its asteroids are at the distance of 40---60 AU from
the Sun.

3. The Oort cloud is located on the Solar system border (more than
100 AU), and its mass estimations are rather vague.

As is known, probability to collide with a heavenly body is in
direct proportion with its surface area. Total mass of the
asteroids is comparable with mass of the Earth, but since there
are so many of them, their overall surface area appears to be
almost equal to the value for the Sun. Besides, there is one more
factor which increases the collision probability. An asteroid mass
is considerably smaller than that of the Sun. This implies that
collision with a PBH should not be necessarily direct to change
the asteroid's trajectory. In order to accomplish this, the PBH
should only be gravitationally scattered on the asteroid, i.e. it
should pass by the asteroid at some distance. This phenomenon
increases the probability for a PBH and an asteroid to interact as
much as for several orders of magnitude.

Thus, PBHs interact much more probably with asteroids than with
the planets or even the Sun.

\section{Probability to find a PBH in the Solar system}

According to our hypothesis we suppose that PBHs are responsible
for almost entire mass of our Galaxy. Simply put, we set the same
mass $m_{pbh}$ for every PBH (in a more complicated case one can
just introduce a distribution function of the PBH mass $m_{pbh}$).

The PBH concentration in the Galaxy is
${n_{pbh}=\rho_{gal}/m_{pbh}}$, where ${\rho_{gal}\approx
10^{-24}}$ g/cm${}^3$ is the average mass density in the Galaxy.

Therefore, the frequency of the Solar system -- PBH collisions is

\begin{equation}
\nu_0 = \pi R_a^2\, v\, \rho_{gal}/m_{pbh} \, , \label{1-1}
\end{equation}
where $R_a$ is the average radius of an asteroid orbit, and  $v$
is the average asteroid velocity in the Galaxy (at infinity with
respect to the Solar system).

\section{Probability for a PBH to collide with an asteroid}

Let the total mass of the asteroids be $M_{ast}$. Density of an
asteroid is usually in the following range: ${\rho_{ast}=2\div 7}$
g/cm${}^3$. Asteroid sizes cover a large interval, with most of
the rocks having less than 1 km in diameter and smaller size
asteroids being more abundant. Since small asteroids do not result
in significant impact, they are not interesting to deal with. So
it makes sense to consider asteroids bigger than some critical
size, e.g. ${D_{ast} \sim 100 m}$. One can estimate their number
as
\begin{equation}
N_{ast} = {\gamma_{ast}\cdot M_{ast}\over \rho_{ast} \cdot
D_{ast}^3 }\, , \label{2-1}
\end{equation}
where $\gamma_{ast}$ is a mass fraction of the asteroids bigger
than $D_{ast}$ in diameter.

The frequency $\nu$ of the PBH -- asteroid interaction is obtained
by multiplying the frequency $\nu_0$ by the probability $\nu_1$
for the PBH to interact with a Solar system asteroid, which is
\begin{equation}
\nu_1=S_{ast}/(\pi R_a^2)\, ,\quad  S_{ast} = N_{ast}\cdot \pi h^2
\, , \label{nu1}
\end{equation}
where $h$ is the maximum impact parameter when the PBH changes
significantly the asteroid trajectory.

Hence, using eq.~(\ref{1-1}) we obtain:
\begin{equation}
\nu = {\pi\, \gamma_{ast}\, \rho_{gal}\, M_{ast}\, v\, h^2 \over
m_{pbh}\, \rho_{ast}\, D^3_{ast} } \, . \label{2-2}
\end{equation}
The quantity $h$ is to be calculated in the next section.

\section{Abrupt change of asteroid orbits due to scattering on PBHs}

With the PBH -- asteroid interaction being the classical two-body
problem, it is more convenient to solve it in the center of mass
frame.

In this reference frame, it is convenient to introduce the
following notation\footnote{Here, $G$ is the gravitation constant,
$M_\odot$ is mass of the Sun,
$M_\oplus$ is mass of the Earth.}:\\
${m=m_{ast}\, m_{pbh}/( m_{ast}+m_{pbh})}$ is the reduced mass, \\
${P=m\, v}$ is the reduced mass momentum in the center of mass frame, \\
${L=m\, v h}$ is the angular momentum of the PBH -- asteroid system, \\
$r_0$ is the minimal distance between the PBH and the asteroid, \\
${U_0=G\, m_{pbh}\, m_{ast}/r_0}$ is an absolute value of the
gravitational interaction energy in the orbit's perihelion, \\
${E={1\over 2}\,  m\, v^2}$ is the total energy of the system, \\
${A = U_0/(2E) = G\, (m_{pbh} + m_{ast})/(r_0 \, v^2)}$ is a
dimensionless scattering parameter, \\
${B=r_0/h}$ is the ratio of the minimal distance to the impact
parameter.

Subtracting the zeroth order $\pi$ and taking into account
relations between the orbit parameters, we obtain the angle of the
momentum deflection after the PBH has been scattered by the
asteroid \cite{1}:
\begin{equation}
\Delta \varphi = 2 \arccos \left[ 1 \right] - 2 \arccos \left[
{-A\over 1+A} \right] -\pi \, . \label{3-1}
\end{equation}
As for B, the calculation yields:
\begin{equation}
B^2 = {1\over 1+2A}\, . \label{3-1-2}
\end{equation}
In the center of mass frame, the change of the momentum is
\footnote{An absolute value of the momentum change is the same
both for the asteroid and the PBH.}:
\begin{equation}
 (\Delta P)^2 = 2P^2 \cdot \left[ 1- \cos(\Delta\varphi) \right] \, .
\label{3-4}
\end{equation}
Using the simple trigonometry and eqs. (\ref{3-1}) and
(\ref{3-4}), we obtain:
\begin{equation}
\Delta P = P\cdot {2 A \over 1+A} \approx 2 P A = {2\, G \, m\,
(m_{pbh} + m_{ast}) \over r_0 \, v} \, . \label{3-3}
\end{equation}

In the Galaxy the PBH average velocity is comparable to the star
velocity dispersion: ${v\approx 300}$~km/s. Therefore, the total
energy $E$ of the system is much more than the gravitational
interaction energy: ${E>>U_0}$, hence, ${A<<1}$, ${B\approx 1}$.

The condition of the crucial alteration of the asteroid orbit is
equality between the initial asteroid momentum,
${P_{ast}=m_{ast}\, v_{ast}}$, and the gained momentum ${\Delta
P}$ within an order of magnitude:
\begin{equation}
\alpha P_{ast} = \alpha\, m_{ast}\, v_{ast} = \Delta P = {2 P A
\over 1+A} \, , \label{3-5}
\end{equation}
where, according to eq. (\ref{3-4}), the parameter ${\alpha
=\sqrt{2(1-\cos\Delta\varphi)}}$ reflects how strongly the
asteroid is affected.

Taking into consideration that the asteroid velocity on a circular
orbit of radius $R$ is ${v_{ast}=\sqrt{G M_\odot /R}}$ and making
use of eq.~(\ref{3-5}), we obtain the exact expression for
$h^{2}$, the mass of asteroid influence having been taken care of:
\begin{equation}
h^2 = {4\, G\, m_{pbh}^2\, R \over \alpha^2 \, v^2 \, M_\odot }
\cdot \left[ 1 - \left( {\alpha (m_{pbh}+m_{ast}) v_{ast} \over 2
m_{pbh} v } \right)^2 \right] \, , \label{3-6}
\end{equation}
or
\begin{equation}
h \approx \left( {m_{pbh}\over \alpha\, M_{\oplus}} \right) \cdot
\sqrt{ {R\over R_{ast}} } \cdot 800\makebox{ km} \, ,
\label{3-6-22}
\end{equation}
where ${R_{ast}\approx 10^{15}}$ cm is the average distance to the
Kuiper belt.

Using this last but one expression, we obtain for $A$:
\begin{equation}
A \approx \alpha \, \sqrt{ {G\, M_\odot \over 4\, v^2\, R }} \cdot
\left( {m_{pbh}+m_{ast} \over m_{pbh} }\right) << 1 \quad
(\makebox{at }\,  m_{pbh} > m_{ast}) \, . \label{3-6-2}
\end{equation}
This last expression validates the above-made assumption on A's
being small for most asteroids of the Kuiper belt.

Eqs. (\ref{nu1}) and (\ref{3-6}) yield the probability for an
asteroid to be deflected by a PBH provided the latter has gotten
into the Solar system:
\begin{equation}
\nu_1\approx\frac{4G m_{pbh}^2\gamma_{ast}M_{ast}}{\alpha^2 v^2
M_\odot R_{ast}\rho_{ast}D_{ast}^3}\approx
\frac{2\cdot\gamma_{ast}\cdot m_{pbh}^2 \cdot (100
\makebox{m})^3}{\alpha^2\cdot M_{\oplus}^2\cdot D_{ast}^3}
\label{nu1-2}
\end{equation}
Substituting eq. (\ref{3-6}) to eq. (\ref{2-2}), we obtain the
frequency of significant orbit deviations for big asteroids:
\begin{equation}
\nu \approx {4\, \pi\, G \, \gamma_{ast}\, \rho_{gal}\, M_{ast}\,
R\, m_{pbh} \over \alpha^2 \, M_\odot  \, \rho_{ast}\, D_{ast}^3
\, v } \, . \label{3-7}
\end{equation}

\section{Conclusion}

It is convenient to rewrite the final expression for the frequency
of the events under consideration as follows:
\begin{equation}
\nu \approx \left(\frac{10^{-7} ys^{-1}}{\alpha^2} \right)\cdot
\left( {m_{pbh}\over M_\oplus} \right)  \cdot \left( {M_{ast}\over
M_\oplus} \right)  \cdot \left({\gamma_{ast}^{1/3}\, 100
\makebox{m}  \over D_{ast}}\right)^3 \cdot \left( {R \over
R_{ast}} \right) \, , \label{4-1}
\end{equation}
The last three factors in formula (\ref{4-1}) are of the order of
unity, so that the final expression only depends on the
distribution and total mass of PBHs in the Galaxy.

Estimating the fraction of asteroids hitting the Earth itself is
quite difficult. This is due to the fact that, even if the
asteroid does not hit some planet, very probably it will not leave
the Solar system. Its orbit now crosses orbits of planets and
other asteroids and becomes more dangerous. So sooner or later the
asteroid itself or secondary asteroids (produced by the very first
one) fall onto the planets.

Thus, the probability for the scattered asteroid to hit the Earth
can be roughly estimated as the ratio of the Earth surface area to
the total surface area of the other planets. It yields
approximately ${\sim 10^{-2}}$. Hence, it does not seem improbable
that formula (\ref{4-1}) may give hit periods of about 190 million
years, which is consistent with observational data.

If the hypothesis analyzed in this paper is correct, modern
methods aimed at averting the asteroid danger appear to be
inefficient. This is related to the fact that their main idea is
revealing big meteors and asteroids with dangerous orbits and,
then, monitoring these bodies. However, if the main danger
consists in abrupt changes of asteroidal orbits (because of
scattering on a PBH), revealing potentially dangerous bodies is
hardly possible.

\section{Acknowledgements}
The online materials from the site http://www.ASTROLAB.ru have
been used by the author while working on this paper.



\begin{thebibliography}{99}

\bibitem{2}
Goncharov, G., and Orlov, V., Astron. Rep. 47, No11, p.925, (2003)

\bibitem{Carr}
Carr, B.J., E-print arXiv: astro-ph/0511743v1, (2005)

\bibitem{1}
Landau, L.D., and Lifshitz, E.M., "Mechanics", Pergamon, Oxford,
(1975)
\end{thebibliography}
\end{document}